\documentclass[preprint]{revtex4}
\usepackage{graphicx,amsmath,amssymb}

\newcommand{\bea}{\begin{eqnarray}}
\newcommand{\eea}{\end{eqnarray}}
\newcommand{\be}{\begin{equation}}
\newcommand{\ee}{\end{equation}}

\begin{document}

\begin{titlepage}
\title{Localization, disorder and boson peak in an amorphous solid}
\author{Leishangthem Premkumar and Shankar P. Das}
\affiliation{School of Physical Sciences,\\
Jawaharlal Nehru University,\\
New Delhi 110067, India.}

\begin{abstract}
We demonstrate using the classical density functional theory (DFT)
model that  an intermediate degree of mass localization in the
amorphous state is essential for producing the boson peak. The
localization length $\ell$ is identified from the width of the
gaussian density profile in terms of which the inhomogeneous density
$n({\bf x})$ of the solid is expressed in DFT. At a fixed average
density, there exists a limiting value $\ell_0$ of $\ell$ signifying
a minimum mass localization in the amorphous state. For more
delocalized states ($\ell>\ell_0$) occurrence of boson peak is
unfeasible .
\end{abstract}

\vspace*{1cm}


\maketitle
\end{titlepage}

Studies of inelastic scattering of light and neutron from amorphous
solids show that the disordered systems are characterized by an
excess density of vibrational modes over that predicted by the
standard Debye distribution. This excess in the density of states
(DOS) $g(\omega)$ for a disordered system typically occurs at the
THz frequency ($\omega$) range and is identified as the so called
boson peak in the $g(\omega)/\omega^2$ vs. $\omega$ plot. The boson
peak height decreases and its location shifts towards the higher
frequencies with the increase of pressure or density
\cite{zeller}-\cite{AM}. A number of different theoretical models
have been developed to explain the anomalous low-energy excitations
\cite{schober}-\cite{dove}. It is now generally believed that the
boson peak is a manifestation of disorder and is key to the
understanding of the vibrational states of glassy materials. The
origin of boson peak in the disordered solid has been often linked
to transverse sound modes for the system\cite{spd_bpk}. It has been
argued that the boson peak results from a band of random transverse
acoustic vibrational states \cite{schirmacher}. Extensive computer
simulations of several glass forming systems \cite{tanaka} also
indicated subsequently that the origin of the boson peak is the
transverse vibrational modes associated with defective soft
structures in the disordered state. In a recent work, Chumakov et
al. \cite{chumakov} explain the boson peak in terms of
(predominantly) transverse sound waves.


A common aspect of a class of the boson peak models mentioned above
is that they are based on the existence of localized modes in the
amorphous solid \cite{zorn}. Longitudinal acoustic phonon modes are
observed in a liquid by ultrasonic, optical, or inelastic scattering
experiments. These hydrodynamic modes occur due to the conservation
laws for the system. For the crystalline state the isotropic
symmetry of the liquid is broken and the constituent particles
vibrate around the sites of a lattice with long range order. This
symmetry breaking leads to the development of the Goldstone modes
\cite{goldstone} or the transverse sound modes in the crystal in
addition to the longitudinal sound modes. For the amorphous solid
the symmetry is not broken over long length scales like that in a
crystal.  When the atomic vibration wavelength in a liquid
approaches the atomic nearest neighbor distance, i.e., in the tera
hertz frequency region, there is a solid like cage formation giving
rise to a restoring force for the acoustic transverse modes.


A typical choice of the order parameter for the crystalline as well
as the amorphous state is the inhomogeneous density $n({\bf r})$
with smooth spatial dependence. The density for the solid state is
expressed in terms of gaussian profiles of width $\ell$ and centered
on the sites of a chosen lattice $\{ {\bf R}_i \}$ \cite{PT}.

\be \label{dens}  n({\bf r})= {\left ( \frac{\alpha}{\pi} \right
)}^{\frac{3}{2}}\sum_{{\bf R}_i} e^{-\alpha {({\bf r}-{\bf
R_i})}^2}, \ee

\noindent with $1/\sqrt{\alpha}=\ell$. The coarse grained picture
for the solid presented above involves two inherent length scales.
First, the microscopic length scale $\sigma$ (say) associated with
the interaction potential between the constituent particles. For a
hard sphere potential, $\sigma$ corresponds to the hard sphere
diameter. Second, the length $\ell$ which represents the degree of
localization of mass in the system. The homogeneous liquid state is
characterized by the limit $\alpha{\sigma^2}\rightarrow{0}$ or
$\ell>>\sigma$.  The sharply localized density profiles of the
crystalline state correspond to the limit $\alpha\sigma^2>> 1$ or
$\ell<<\sigma$. The ratio $\ell/\sigma$ is generally referred to as
the Lindeman parameter for the crystalline. In the present work we
compare the free energy contribution due to the vibrational modes in
the amorphous state to the corresponding density functional result
for the ideal gas part $f_\mathrm{id}[n]$\cite{spd_book} of the free
energy for an inhomogeneous density $n({\bf r})$. We show that for a
a chosen average density, the characteristic gaussian profile of
$n({\bf r})$ signifies a minimum localization length $\ell_0$ below
which the appearance of boson peak in the reduced density of states
is unfeasible.

The equilibrium state of the fluid is identified by minimizing a
proper thermodynamic potential or free energy with respect to the
density $n({\bf x})$. The free energy is expressed as a sum of two
parts which are respectively the free energy of the non-interacting
system $F_\mathrm{id}$ and the contribution $F_\mathrm{ex}$ due to
interactions between the particles. $F_\mathrm{id}$ is obtained from
the logarithm of the partition function ${(V/\wedge^3)}^N/N!$ for
the noninteracting system of $N$ particle, where
$\wedge=\hbar/\sqrt{2\pi{m}k_BT}$ denotes the thermal wave length at
temperature $T$. Thus ${\beta}F_\mathrm{id}=Vn_0(\ln({n_0}\wedge^3)
- 1)$ for the uniform density fluid $n_0$. For the inhomogeneous
density $n({\bf x})$, the expression for the free energy is easily
generalized to $\beta {F}_\mathrm{id}[n] = \int d {\bf r} n({\bf r})
[\ln\{n({\bf r})\wedge^3 \}-1]$. For the uniform density $n_0$
changing to a nonuniform one $n({\bf x})$, the ideal gas
contribution for the $N$ particles system changes by an amount
$n_0\int d{\bf r} \tilde{n}({\bf r}) [\ln \tilde{n}({\bf r}) ]$,
where $\tilde{n}({\bf r})=n({\bf r})/n_0$. Since by definition
$\tilde{n}({\bf r})$ is always positive, using the Gibbs inequality
$\{x\ln{x} - x + 1\}\geq 0$ for positive $x$, it follows that
entropy drops upon localization of the particles. This is a result
of the restriction of available phase space. If $\alpha$ is larger
than a limiting value $\alpha_0$ (say) the density $n({\bf r})$ is
well approximated in terms of the gaussian profile centered at the
nearest lattice site. The free energy per particle $f_\mathrm{id}$
for large $\alpha\sigma^2$ is obtained as,

\begin{equation}
\label{dftapp} f_\mathrm{id}[\alpha] \approx -\frac{5}{2} + 3\ln
\left ( \sqrt{\frac{\alpha}{\pi}}\wedge \right ).
\end{equation}

In a microscopic approach, vibrational modes with localized density
profiles contribute to the free energy and the latter is obtained in
terms of the density of vibrational states $g(\omega)$ of the
system. The free energy of the non-interacting system is obtained as
$F_\mathrm{id}=-k_{B}T\ln Z_N~$ where $Z_N$ is the partition
function for 3N harmonic oscillators in constant NVT ensemble. Going
from discrete to continuum in the frequency spectrum, we obtain the
ideal gas free energy per particle in units of $\beta^{-1}$ as
${\beta}F_\mathrm{id}/N  = \int_0^1 \kappa(x_mx) \bar{g}(x) dx
~~\equiv f_0 [ \bar{g},x_m(\alpha)]$, where
$\kappa(x)=-\frac{1}{4}\{x+\ln[1-e^{-x}]-2x/(e^x-1)\}$. $\omega_m$
is the upper cutoff of frequency up to which $g(\omega)$ is nonzero
and it is expressed in terms of the dimensionless quantity
$\beta\hbar{\omega}_m=x_m {\equiv} x_m(\alpha)$. The scaled density
of states $\bar{g}(x)$ is obtained from $g(\omega)$ as
$\bar{g}(x)={(3n_0)}^{-1}\omega_m g(\omega)$ where
$x=\omega/\omega_m$ is the reduced frequency.  The expression for
$\kappa(x)$ indicates the functional dependence of the integral
$f_0$ on the scaled density of states $\bar{g}$. There are $3n_0$
vibrational modes for an average $n_0$ number of particles in a unit
volume giving rise to the normalization condition
$\int_0^{\omega_m}g(\omega){d\omega}=3n_0$ for $g(\omega)$. The
cutoff frequency $\omega_m$ is a characteristic property of the
material and sets the shortest length scale for the vibrational
modes. For the Debye distribution, $g(\omega)=g_D(\omega) \equiv
(9n_0/\omega_D^3)\omega^2$ is nonzero in the range
$0\leq{\omega}\leq{\omega}_D$. The Debye frequency $\omega_D$ is
obtained as $\omega_D={(6\pi{n_0})}^{1/3}c$, with $c$ being the
speed of sound. The scaled (Debye) distribution $\bar{g}_D$ as
$\bar{g}_D(x)=3x^2$ where  $x=\omega/\omega_D$. Considering the case
$x_m{\equiv}x_D=\beta\hbar\omega_D << 1$, analysis of the integral
$f_0 [ \bar{g},x_m(\alpha)]$ for $\bar{g}(x)\equiv{\bar{g}_D(x)}$
obtains the following result to leading orders in $x_D$:

\begin{equation}
\label{etsasym} f_0 [\bar{g}_D,x_D] =-\frac{5}{2}+3\ln x_D+O(x_D^2).
\end{equation}

\noindent This mapping identifies the Debye cutoff frequency
$\omega_D$ in terms of the corresponding width parameter $\alpha$ of
density functional description as $\beta\hbar\omega_D {\equiv}
x_D(\alpha) =\sqrt{{\alpha}/{\pi}}\wedge$. For a given $\wedge$, a
corresponding $\alpha_0$ is identified such that for all
$\alpha\geq{\alpha}_0$ the asymptotic form (\ref{dftapp}) will
represent the ideal gas free energy. The value of $\wedge/\sigma$
should be such that $x_D(\alpha_0)\leq{x_L}$ where the asymptotic
form (\ref{etsasym}) for $f_0(x)$ holds for $x<{x_L}$.


In DFT the free energy $f_\mathrm{id}(\alpha)$ for the amorphous
state is evaluated numerically using the $n({\bf r})$ corresponding
to the parameter value $\alpha$. The centers for the gaussian
density profiles $\{{\bf R}_i\}$ are distributed on a random lattice
and the corresponding site-site correlation function $g(R)$ is
obtained from the Bernal's \cite{Bernel} random structure generated
through the Bennett algorithm \cite{Benett}. We approximate the
$g(R)$ through the following relation $g(R)=g_B[ R
{(\eta/\eta_0)}^{1/3}]$ where $\eta$ denotes the average packing
fraction and $\eta_0$ is used as a scaling parameter for the
structure \cite{lowen} such that at $\eta=\eta_0$ Bernal's structure
is obtained. Smaller the parameter $\alpha$ in $n({\bf x})$, larger
number of sites are needed to be included for an accurate evaluation
of the $f_\mathrm{id}$. The free energy evaluated numerically agrees
with the corresponding asymptotic expression (\ref{dftapp}) for
$\alpha\geq\alpha_0$. This is shown in the inset of Fig. 1. The
difference $\Delta{f}_\mathrm{id}$ between the two evaluations of
$f_\mathrm{id}$ is shown in Fig. 1 for $\alpha<\alpha_0$ range. This
result corresponds to hard sphere system with the fixed average
density $n_0\sigma^3=1.1$ or packing fraction
$\varphi=\pi{n_0}\sigma^3/6=.576$. The thermal wavelength is kept
constant at $\wedge/\sigma=0.0137$. The free energy obtained from
the DFT expression $f_\mathrm{id}[n({\bf r})]$ corresponding to
$n({\bf r})$ in the range $\alpha<\alpha_0$, deviates completely
from the asymptotic result (\ref{dftapp}). It extends to the correct
$\alpha{\rightarrow}0$ limit $(\ln\{n_0\wedge^3\}-1)$ for the
uniform liquid state. A key observation here is that the free energy
curve for small $\alpha<\alpha_0$ matches with the corresponding
result obtained from the microscopic expression $f_0
[\bar{g},x_m(\alpha)]$ with a modified density of states which is
different from the Debye distribution. The normalization condition
for $g(\omega)$ is maintained with a corresponding
$\omega_m(\alpha)$ or equivalently $x_m(\alpha)$ different from that
for the Debye case ($=\sqrt{{\alpha}/{\pi}}\wedge$). Therefore we
require that the function $x_m(\alpha)$ $\rightarrow$ $x_D(\alpha)$
as $\alpha\rightarrow{\alpha_0}$. The correction part in this
functional relation is denoted by $C(\alpha)$, such that
$x_m(\alpha)=x_D(\alpha_0)+C(\alpha)$ with $C(\alpha){\rightarrow}0$
as $\alpha{\rightarrow}\alpha_0$. Similarly, for the scaled
distribution function $\bar{g}(x)$ we use a correction over the
corresponding Debye form $\bar{g}_D(x)=3x^2$ in terms of the
function $\Delta$, such that $\bar{g}(x,x_m(\alpha))=
3x^2(1+\Delta(x,\alpha))$. Since for all frequencies $x$, we must
have $g\rightarrow{\bar{g}_D}$ as $\alpha{\rightarrow}\alpha_0$, we
express $\Delta$ with the separation of variables. Thus
$\Delta(x,\alpha)=B(\alpha)\tilde{\Delta}(x)$ where
$B(\alpha){\rightarrow}0$ as $\alpha{\rightarrow}\alpha_0$.
Furthermore in order to maintain the normalization condition of
$g(\omega)$, the function $\tilde{\Delta}(x)$ for all $\alpha$ must
satisfy $\int_0^1 x^2 \tilde{\Delta}(x)dx=0$. The two functions
$C(\alpha)$ and $B(\alpha)$ are parameterized in terms of two
respective polynomials of the separation parameter
$\epsilon=1-\alpha/\alpha_0$. These functions are determined by
requiring that the $f_0 [\bar{g},x_m(\alpha)]$  obtained with the
corresponding $\bar{g}$ and $x_m$ agree with that of the DFT
expression for $f_\mathrm{id}$ in the range $\alpha{\leq}\alpha_0$.
We choose the $x$ dependence of the $\tilde{\Delta}(x)$ so as to
have an intermediate peak over the whole frequency range. The
relative frequency $x$ at the peak of $\tilde{\Delta}(x)$ is a
parameter which is kept fixed through out this work. In order to
maintain the positivity of the density of states the function
$\tilde{\Delta}(x)$ is also kept confined in the range $\pm{1}$. In
Fig. 2 we show the results obtained for the appropriate density of
states $\bar{g}(x)$ which satisfy the above constraints of density
functional theory. The reduced density of states
$g(\omega)/\omega^2$ vs. $\omega/\omega_D$  is shown in Fig. 2 for
three different values of the localization parameter
$\ell=1/\sqrt{\alpha}$. The height $h_B$ of the boson peak increases
as $\ell$ grows and is shown in the inset of Fig. 2. Larger $\ell$
or smaller $\alpha$ corresponds to increased spreading of the
density profiles and hence decreased localization.

As the density profiles get increasingly spread out, {\em i.e.}, as
$\alpha$ decreases, the intensity or height of the boson peak
continues to grow while upper cutoff $\omega_m$ of the frequency
keeps decreasing. However, this trend does not continue indefinitely
towards $\alpha{\rightarrow}0$. To understand this behavior we note
that the upper cutoff $\omega_m$ sets the shortest wavelength
$\lambda_\mathrm{min}$ for the sound waves. In a valid continuum
description of the solid, $\lambda_\mathrm{min}$ cannot get smaller
than the average mean square displacement of the individual
particles. In the density functional theory formulated in terms of
gaussian density profiles, the average particle displacement is
$\ell=1/\sqrt{\alpha}$ and this is a reasonable estimate for
$\lambda_\mathrm{min}$. The corresponding upper limit in the
frequency of the sound wave is obtained from the relation
${\Omega^{L}}/(2\pi{c})=\sqrt{\alpha}$ where $c$ is the speed of
sound.  In Fig. 3, for $n_0\sigma^3=1.1$ this limiting value for the
upper limit vs. cutoff $\sqrt{\alpha}$ is shown as a dashed line.
The corresponding upper cutoff of the frequency $\omega_m(\alpha)$
obtained by matching microscopic expression for the free energy with
the density functional expression (as described with respect to Fig.
2) is shown as a solid line. We mark with an arrow in Fig. 3, the
point $\alpha=\alpha_\mathrm{min}$ where the two lines cross. This
point depicts the maximum delocalized state for which the boson peak
in $g(\omega)$ is observed in the vibrational density of states. For
more delocalized states the description in terms of sound waves
breaks down. In the inset of Fig. 3, we show how the point of
maximum delocalization obtained in the main figure changes with the
density of a hard sphere system.


Metastable states of strongly heterogeneous density distribution has
been obtained \cite{spd_kaur,kim-munakata,chandan-sood} with free
energy minimum intermediate between the uniform liquid and crystals
with long range order. We regulate the parameter $\eta_0$ to define
specific sets of random structures. The low $\alpha$ or partially
delocalized state exists only if total free energy is computed with
the $n({\bf x})$ for lattice points $\{ {\bf R}_i \}$ on a random
structure. Thus disorder is essential for locating the metastable
minimum of the free energy in the delocalized region
($\alpha\leq{\alpha}_0$). This is due to the interaction
contribution to the free energy. However the corresponding
$F_\mathrm{id}$ in the small $\alpha$ region remains the same with
$\{ {\bf R}_i \}$ representing a regular crystalline lattice with
long range order instead of the Bernel structure. With
$m_i=\exp[-K_i^2/(4\alpha)]$ the definition (\ref{dens}) for the
density $n({\bf r})$ reduces to the form $n({\bf r})=n_0+\sum_i m_i
e^{i{\bf K}_i.{\bf r}}$  in terms of an expansion\cite{shenoxtoby}
involving reciprocal lattice vector (RLV) $\{{\bf K}_i\}$ .  It is
useful to note at this point that the boson peak in disordered
system has been studied in terms of a geometrically perfect crystal
having random interactions between the neighbors
\cite{schirmacher,Elliott}.

For the hard sphere system with $n_0\sigma^3=1.1$ we obtain the
interaction contribution to the free energy in terms of standard
Ramakrishnan-Yussouff functional \cite{RY} using the solution of the
Percus-Yevick equation with Verlet-Weiss\cite{HG}. For the
heterogeneous glassy state, the free energy barrier $f_B$ is
obtained as the height of the maximum in the corresponding free
energy curve from the metastable minimum at small nonzero $\alpha$
(inset (a) of Fig. 4). For the optimum value of $\alpha$ signifying
a metastable state, $g(\omega)$ is obtained. In Fig. 4 we show the
variation of $f_B$ for the parameter range $.62<\eta_0<.68$, with
the corresponding boson peak height $h_B$. As the free energy
barrier $f_B$ becoming weaker, the height of the boson peak $h_B$
goes down. In the inset (b) of Fig. 4 is shown the $f_B$ vs.
$\ell=1/\sqrt{\alpha}$ curve. As the mass localization gets weaker,
{\em i.e.} with decreasing(increasing) of $\alpha$($\ell$) the
barrier height $f_B$ gets smaller. A lower barrier height implies
that the structural degradation is less hindered in the
corresponding system, {\em i.e.}, represents a glassy material of
higher fragility. The dependence of the peak height $h_B$ and the
peak frequency $\omega_p$ on the pressure are respectively shown in
Fig. 5 and its inset and conforms to experimental findings.

An important characteristic of the boson peak is that it gets
weaker, more fragile the system is. This link between the long-time
relaxation behavior (fragility) of the glass forming material and
the vibrational modes in the THz region, follows in a natural manner
in the present model. Since less fragile or stronger liquids tend to
form network structures, the density profiles are more localized for
them. Therefore increase of fragility is synonymous with decrease of
$\alpha$, {\em i.e.}, less localized density profiles and lower
barriers. Since as discussed above the boson peak height decreases
with decreasing $\alpha$, it also implies weaker boson peak for more
fragile systems \cite{sokolov,angell}. Our present approach of DFT
is similar to the models of disordered solids in terms of springs.
For large values of $\alpha$, the sharply localized density profiles
are interpreted in terms of harmonic oscillators with spring
constant $\kappa$ which is proportional to the width parameter
$\alpha$ \cite{kirkwolynes}. A natural extension of the present
model will be to include fluctuations of the widths $\alpha$ at
different sites on the amorphous structure as an appropriate
description of the heterogeneous glassy state. PK acknowledges CSIR,
India for financial support. SPD acknowledges BRNS, India for
financial support under 2011/37P/47/BRNS.

\newpage
\begin{figure}[!htb] 
\centering
\includegraphics[width=10cm]{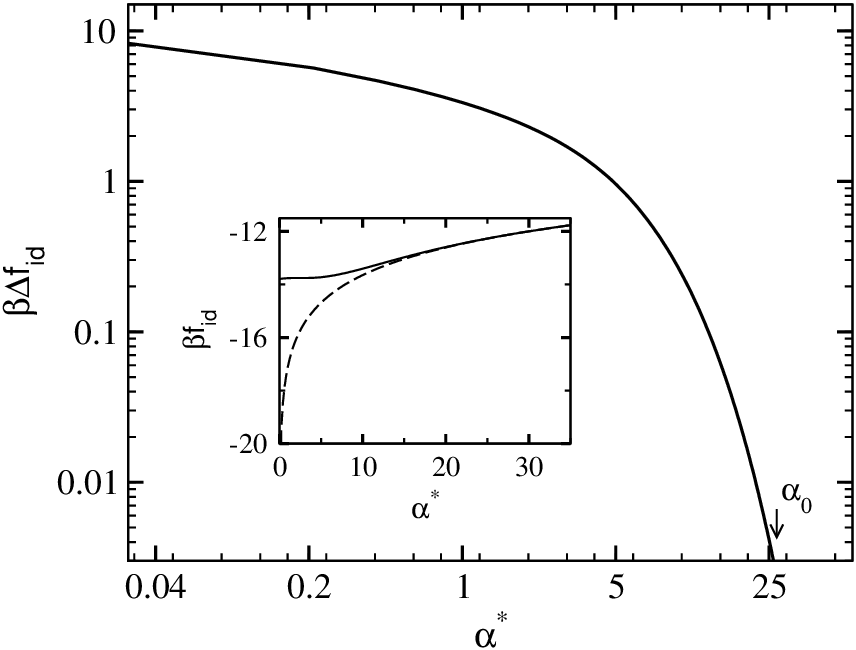} 
\caption{Inset : ideal gas part of free energy  $f_\mathrm{id}$ per particle
from numerical evaluation of the density functional expression (solid line) and
from the asymptotic formula (\ref{dftapp}) (dashed line). The difference between the
the asymptotic formula and the numerical result for $f_\mathrm{id}$
is shown in an enlarged form in the main figure for the $0<\alpha<\alpha_0$ region.}
\label{fid-fig}
\end{figure}

\begin{figure}[!htb] 
\centering
\includegraphics[width=10cm]{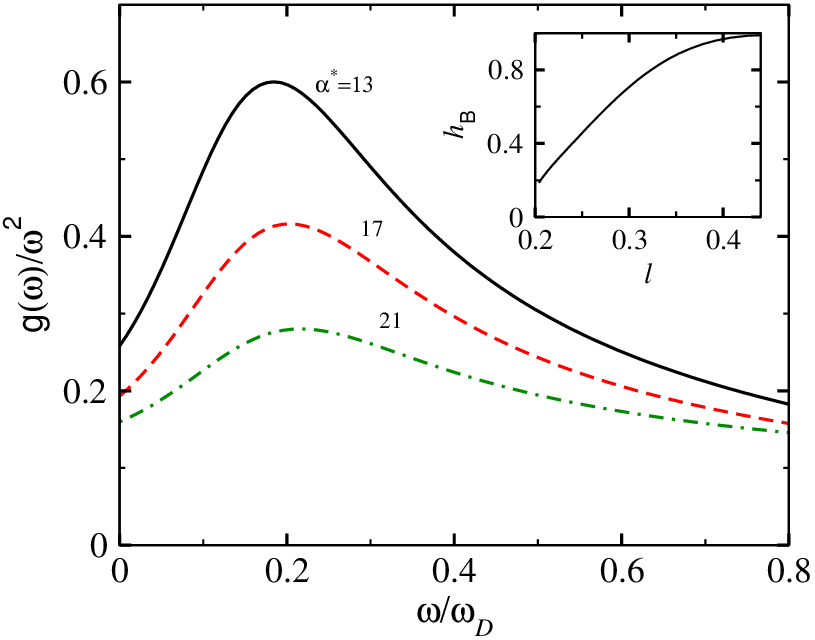} 
\caption{The reduced density of states $g(\omega)/\omega^2$ in units
of $c^3$ ( $c$ is the sound speed) vs. $\omega/\omega_D$ corresponding to
$\alpha\sigma^2$: 13 (solid), 17(dashed), 21(dot-dashed),
for a hard sphere system at density of $\rho_0\sigma^3=1.1$.
Inset shows corresponding boson
peak height $h_B$ vs. localization length $\ell$
($=1/\sqrt{\alpha}$).}
\end{figure}

\begin{figure}[!htb] 
\centering
\includegraphics[width=10cm]{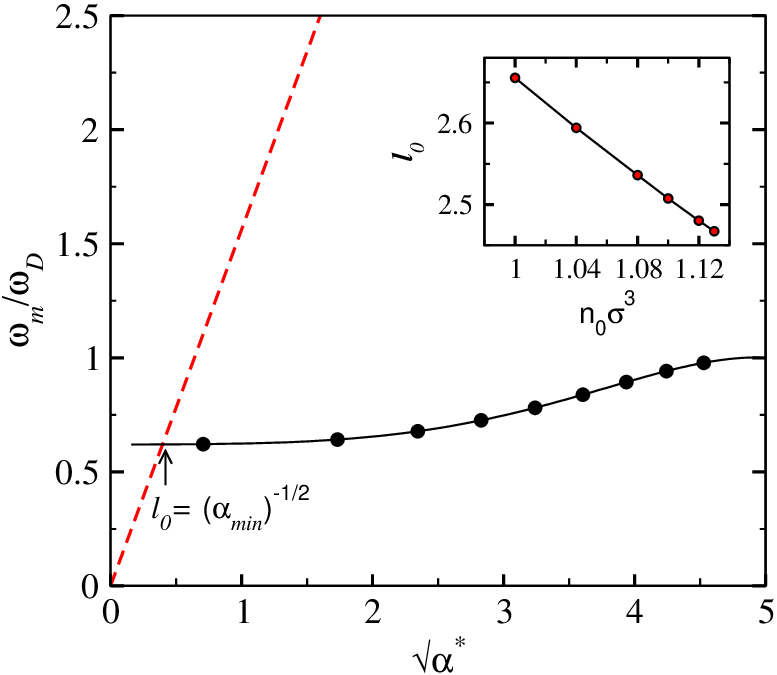} 
\caption{For a hard sphere system at density $\rho_0\sigma^3=1.1$,
the ratio  $\omega_m/\omega_D$ (see text) vs. $\sqrt{\alpha}\sigma$
(filled dots). Dashed line shows the upper limit of the frequency of
the sound waves for localized the vibrational modes (see text).
$\alpha_\mathrm{min}$ shown with an arrow is the minimum possible
value of $\omega_m/\omega_D$ for the boson peak. Inset shows
variation of $\ell_0$ ($=1/\sqrt{\alpha_\mathrm{min}}$ in units of
$\sigma$ ) with density $\rho_0\sigma^3$ for the hard sphere
system.}
\end{figure}

\begin{figure}[!htb] 
\centering
\includegraphics[width=10cm]{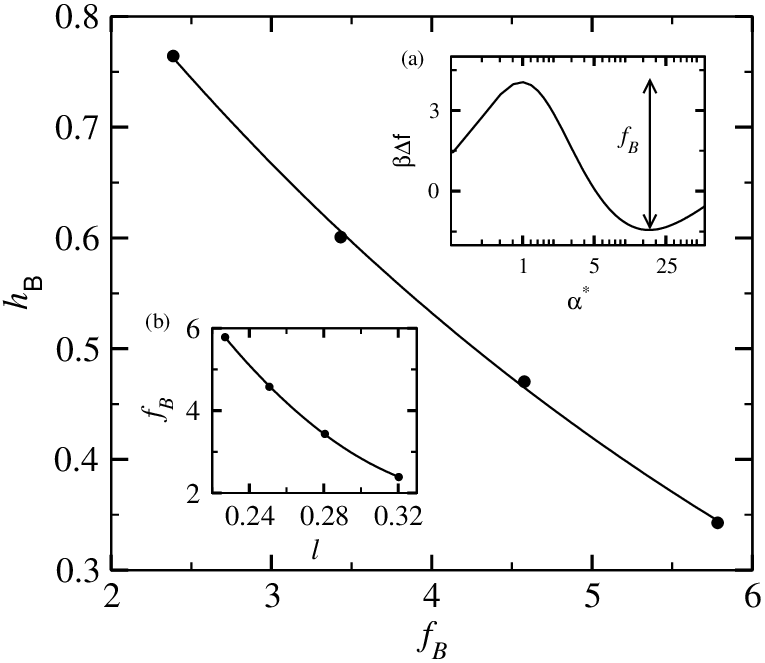} 
\caption{Boson peak height $h_B$ vs. the free energy barrier $f_B$
for the heterogeneous metastable states obtained in the
$\alpha_0>\alpha$ region corresponding to different structures of a
hard sphere system of $\rho_0\sigma^3=1.1$. Inset a) definition of
$f_B$ and b) $f_B$ vs. the width $\ell$ of the gaussian density
profiles.}
\end{figure}

\begin{figure}[!htb] 
\centering
\includegraphics[width=10cm]{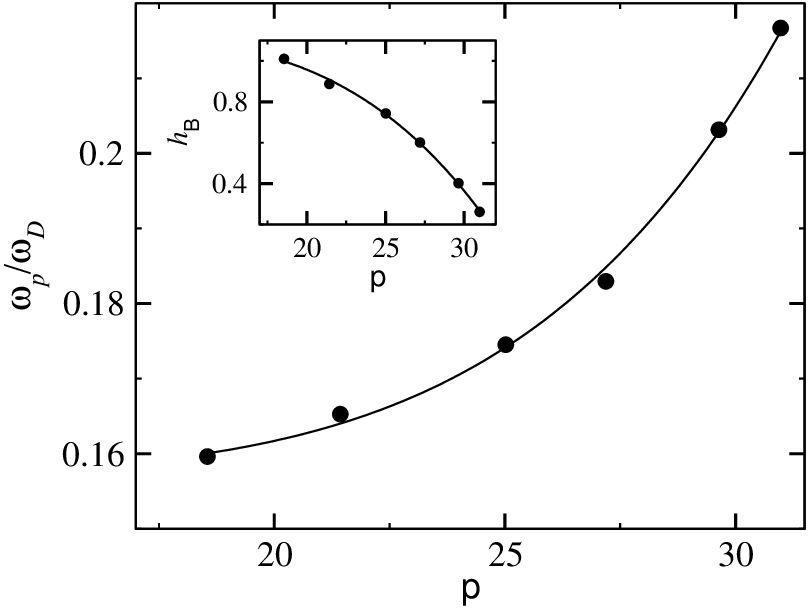} 
\caption{Ratio of frequency $\omega_p$ at the boson peak to the
Debye frequency $\omega_D$ vs. pressure $P$ of the system ( in units
of ${(\beta \sigma^3)}^{-1}$). Inset shows height $h_B$ of the peak
vs. pressure $P$.}
\end{figure}

\end{document}